\documentclass[onecolumn, notitlepage,secnumarabic, amssymb, nobibnotes, aps, pra]{revtex4-1}
\usepackage{graphicx}
\usepackage{color}
\usepackage{amsmath,amsfonts,amssymb,amsthm, bm}

\begin{document}

%\title{Derivation of Frequency Comb Measurement using Heterodyne technique}
\title{Heterodyne measurement of sidebands and frequency combs: a derivation}

\author{Rory W. Speirs}
\email[]{speirs@umd.edu}
\affiliation{Joint Quantum Institute, National Institute of Standards and Technology and the University of Maryland, College Park, MD 20742, USA}

%\date{\today}
\date{November 2019}

\begin{abstract}
A mathematical description of heterodyne measurement of an optical frequency comb is presented. It is shown that for a signal beam containing many frequency teeth, the amplitude and phase of each tooth can be determined from the beatnote generated when the signal is interfered with a local oscillator with known offset frequency.
\end{abstract}

\maketitle

\section{Introduction}
Optical heterodyne detection is a commonly used method for measurement of amplitude and phase of spectral components of an optical beam \cite{kuri2003optical, delange1968optical, chtcherbakov2007optical, hall1981optical, jacobs1988optical}. Despite its frequent usage, complete and concise mathematical descriptions of the heterodyne measurement process are difficult to come by, with many sources simply stating a non-generalised final result \cite{keyes2013optical, menzies2005laser}. The lack of a readily available comprehensive description can make it difficult to deeply understand the method, including underlying assumptions, which increases the chance of experimental or analytical errors.

Here, we present a complete and concise derivation of optical heterodyne measurement of a beam consisting of regularly spaced frequency components. Such beams are commonly generated using electro-optic modulators, and ultrafast laser techniques, and where the number components is large, the optical field is referred to as a frequency comb. We attempt to use consistant notation, and where appropriate, give a physical interpretation to the mathematical results. 

At a given position any electric field strength can be written as
\begin{equation}
	E(t) = \frac{1}{2}\mathcal{E}(t)e^{-{\rm{i}}\omega_0t} + \frac{1}{2}\mathcal{E}^*(t)e^{{\rm{i}}\omega_0t}
	\label{a}
\end{equation}
where $\mathcal{E}(t)$ is the complex envelope function, $\mathcal{E}^*$ is its complex conjugate, and $\omega_0$ is the reference angular frequency. The choice of reference frequency is completely arbitrary, and can be positive or negative. However, the reference is \emph{usually} chosen so that the variation in time of $\mathcal{E}(t)$ is as small as possible. In practice, the reference frequency is often taken as the carrier frequency of a beam, so the complex envelope fully represents only the modulations to this carrier (both in phase, amplitude, and any additional frequency components).

The intensity of a given electric field is given by
\begin{equation}
	I(t) = c\varepsilon \frac{1}{T}\int_{-T/2}^{T/2} E^*(t)E(t){\rm{d}}t,
	\label{ba}
\end{equation}
where $T$ is an integer number of periods of the optical cycle, with minimum one period.

Subbing equation \ref{a} into this expression yields:
\begin{align}
	I(t) &= \frac{c\varepsilon}{4T}\int_{-T/2}^{T/2} \left[\mathcal{E}^*(t)\mathcal{E}(t) + \mathcal{E}(t)\mathcal{E}^*(t) + \mathcal{E}^2(t)e^{-{\rm{i}}2\omega_0t} + \mathcal{E}^{*2}(t)e^{{\rm{i}}2\omega_0t}  \right]{\rm{d}}t \label{b0}\\
	     & = \frac{c\varepsilon}{2T}\int_{-T/2}^{T/2} \mathcal{E}^*(t)\mathcal{E}(t){\rm{d}}t + \frac{c\varepsilon}{4T}\int_{-T/2}^{T/2} \left[ \mathcal{E}^2(t)e^{-{\rm{i}}2\omega_0t} + \mathcal{E}^{*2}(t)e^{{\rm{i}}2\omega_0t}  \right]{\rm{d}}t
	\label{b}
\end{align}
If we have chosen $\omega_0$ appropriately such that $\mathcal{E}$ is slowly varying in time, the expression in the first integral can be pulled out of the integral, and the integral evaluates to $T$. The two terms on the right oscillate at twice the optical frequency, and so will integrate to zero. Note this definition of intensity breaks down where the amplitudes vary significantly over the time scale of an optical cycle, such as with ultra-short laser pulses.

Averaging over the fast oscillations gives a convenient expression for intensity written in terms of the complex amplitudes:
\begin{align}
	I(t) &= \frac{c\varepsilon}{2}\mathcal{E}^*(t)\mathcal{E}(t)\label{c0}\\
	     &=\frac{c\varepsilon}{2}|\mathcal{E}(t)|^2.
	\label{c}
\end{align}

Now, when we interfere two beams, we will have the signal $E_{\rm{sig}}(t)$ and the local oscillator $E_{\rm{LO}}(t)$ fields.
The signal can be written as
\begin{equation}
	E_{\rm{sig}}(t) = \frac{1}{2}\mathcal{E}_{\rm{sig}}(t)e^{-{\rm{i}}\omega_0t} + \frac{1}{2}\mathcal{E}_{\rm{sig}}^*(t)e^{{\rm{i}}\omega_0t}.
	\label{d}
\end{equation}

A frequency comb with tooth spacing $\omega_{\rm{SB}}$ (`SB' is signifies `side band'), is written as a sum of fields each with a constant amplitude and a frequency that differs by an integer multiple of $\omega_{\rm{SB}}$. For compactness, a reference frequency can be pulled out of each term, and the total signal field can be written as equation \ref{d}, where the complex envelope is given by:
\begin{equation}
	\mathcal{E}_{\rm{sig}}(t) = \sum_{n}a_{n}e^{-{\rm{i}}(n\omega_{\rm{SB}}t + \phi_{n})}
	\label{e}
\end{equation}
where $n$ can take any integer value (positive, negative or zero), $a_n$ is the real amplitude of the $n^{\rm{th}}$ comb tooth, and $\phi_{n}$ is the corresponding phase of that frequency component (cf. \ref{apxc1}). Note that this is simply a general expression that represents a frequency comb, it is not a recipe for how to generate one. It is well known that modulating either the phase or amplitude of a carrier periodically puts sidebands on that carrier, but here we are simply concerned with how to represent (and then measure) a field that \emph{does} have evenly spaced sidebands.

The full signal electric field is therefore written as:
\begin{equation}
	E_{\rm{sig}}(t) = \frac{1}{2}\sum_{n}a_{n}\left[e^{-{\rm{i}}(n\omega_{\rm{SB}}t + \phi_{n})}e^{-{\rm{i}}\omega_0t} +e^{{\rm{i}}(n\omega_{\rm{SB}}t + \phi_{n})}e^{{\rm{i}}\omega_0t}\right].
	\label{f}
\end{equation}

In a heterodyne measurement, the signal beam is combined with a local oscillator beam, which is shifted in angular frequency relative to the signal carrier by an amount $\Delta\omega$. The complex envelope can therefore be written as:
\begin{equation}
	\mathcal{E}_{\rm{LO}}(t) = be^{-{\rm{i}}(\Delta\omega t + \phi_{\rm LO})},
	\label{g}
\end{equation}
where $b$ is the local oscillator real amplitude.

The full local oscillator electric field can then be written as:
\begin{equation}
	E_{\rm{LO}}(t) = \frac{b}{2}\left[e^{-{\rm{i}}(\Delta\omega t + \phi_{\rm LO})}e^{-{\rm{i}}\omega_0t} + e^{{\rm{i}}(\Delta\omega t + \phi_{\rm LO})}e^{{\rm{i}}\omega_0t}\right].
	\label{h}
\end{equation}

Now, when the signal and local oscillator beams are combines on a beamsplitter, the output of a single port is just the sum of their fields. We will only be interested in a single port, so the phase shift introduced by reflection of a mirror is irrelevant to us and will be ignored (as will the halving of their amplitudes). The result is:
\begin{align}
	E_{\rm{tot}}(t) &= E_{\rm{sig}}(t) + E_{\rm{LO}}(t)\\
	     &= \frac{1}{2}\left[\sum_{n} a_{n}e^{-{\rm{i}}(n\omega_{\rm{SB}}t + \phi_{n})} + be^{-{\rm{i}}(\Delta\omega t + \phi_{\rm LO})} \right]e^{-{\rm{i}}\omega_0t}
	        +\frac{1}{2}\left[\sum_{n} a_{n}e^{{\rm{i}}(n\omega_{\rm{SB}}t + \phi_{n})} + be^{{\rm{i}}(\Delta\omega t + \phi_{\rm LO})} \right]e^{{\rm{i}}\omega_0t}
	\label{i}
\end{align}
where the terms have been grouped by pulling out their common exponential factor.

We can see that equation \ref{i} is in the form of equation \ref{a}, where the complex envelope function is given by the term in the square brackets. We can therefore immediately write down the resulting time varying intensity for the combined field by substituting the envelope and its conjugate into equation \ref{c0}. Note that when calculation the intensity the summations over the signal teeth will multiply and hence the summations must be performed separately. The index label of one summation is therefore changed to $m$.

\begin{align}
	I_{\rm{tot}}(t) &= 	\frac{c\varepsilon}{2} \left[\sum_{n} a_{n}e^{-{\rm{i}}(n\omega_{\rm{SB}}t + \phi_{n})} + be^{-{\rm{i}}(\Delta\omega t + \phi_{\rm LO})} \right] \times 
					\left[\sum_{m}a_{m}e^{{\rm{i}}(m\omega_{\rm{SB}}t + \phi_{m})} + be^{{\rm{i}}(\Delta\omega t + \phi_{\rm LO})} \right] \label{j0}\\
	     &=  \frac{c\varepsilon}{2}\left\{ 
	     b^2 + \sum_{n, m} a_{n}a_{m}e^{{\rm{i}}\left[(m-n)\omega_{\rm{SB}}t + \phi_{m} - \phi_{n}\right]}
	      + \sum_{m}a_{m}be^{{\rm{i}}\left[(m\omega_{\rm{SB}}-\Delta\omega)t + \phi_{m} - \phi_{\rm LO}\right]}
	      + \sum_{n}a_{n}be^{{\rm{-i}}\left[(n\omega_{\rm{SB}}-\Delta\omega)t + \phi_{n} - \phi_{\rm LO}\right]}
	     \right\}\\
	     &=  \frac{c\varepsilon}{2}\left\{ 
	     b^2 + \sum_{n}a_{n}^2 
	     + \sum_{\substack{n,m\\ n\neq m}} a_{n}a_{m}e^{{\rm{i}}\left[(m-n)\omega_{\rm{SB}}t + \phi_{m} - \phi_{n}\right]}
	      + \sum_{n}a_{n}be^{{\rm{i}}\left[(n\omega_{\rm{SB}}-\Delta\omega)t + \phi_{n} - \phi_{\rm LO}\right]}
	      + \sum_{n}a_{n}be^{{\rm{-i}}\left[(n\omega_{\rm{SB}}-\Delta\omega)t + \phi_{n} - \phi_{\rm LO}\right]}
	     \right\}
	      \label{j1}
\end{align}

So, equation \ref{j1} tells us exactly what the \emph{intensity} of the combined field will be at a given point in time. Clearly the intensity will vary periodically in time so this is commonly referred to as the \emph{beat note}, although in the case of a signal beam which is made up of many comb teeth, there will be multiple frequency components present.

We are often primarily interested in knowing something about these different frequency components, so a common way to analyse the heterodyne signal is using a spectrum analyser to display at the power spectrum of the photocurrent from a photodiode. The photocurrent is directly proportional to intensity, so the generated power spectrum is directly proportional to the power spectrum of the intensity. A spectrum analyser simply Fourier transforms the signal, and then displays the amplitude (or power, which is proportional to the square of the amplitude), so it is useful to see what the analytic form is of the Fourier transform of the time varying intensity, $\breve{I}(\omega)$:
\begin{equation}
	\breve{I}(\omega) = \frac{1}{\sqrt{2\pi}}\int I(t)e^{{\rm{-i}}\omega t}{\rm{d}}t,
	\label{k}
\end{equation}
where the integral is understood to be over $-\infty$ to $\infty$.

Substituting equation \ref{j1} into the Fourier transform yields:

\begin{align}
	\breve{I}_{\rm tot}(\omega) &=    \frac{c\varepsilon}{2\sqrt{2\pi}}  \int        
	    \Big\{
	     b^2 + \sum_{n}a_{n}^2 
	     + \sum_{\substack{n,m\\ n\neq m}} a_{n}a_{m}e^{{\rm{i}}\left[(m-n)\omega_{\rm{SB}}t + \phi_{m} - \phi_{n}\right]} \notag\\
	      &\qquad\qquad+ \sum_{n}a_{n}be^{{\rm{i}}\left[(n\omega_{\rm{SB}}-\Delta\omega)t + \phi_{n} - \phi_{\rm LO}\right]}
	      + \sum_{n}a_{n}be^{{\rm{-i}}\left[(n\omega_{\rm{SB}}-\Delta\omega)t + \phi_{n} - \phi_{\rm LO}\right]}
	     \Big\}
	     e^{{\rm{-i}}\omega t}{\rm{d}}t  \label{l0}\\
	     &=  \frac{c\varepsilon}{2\sqrt{2\pi}}         
	    \Big\{
	     \left(b^2 + \sum_{n}a_{n}^2\right)\int e^{{\rm{-i}}\omega t}{\rm{d}}t
	     + \sum_{\substack{n,m\\ n\neq m}} a_{n}a_{m}\int e^{{\rm{i}}\left[(m-n)\omega_{\rm{SB}}t + \phi_{m} - \phi_{n}\right]}e^{{\rm{-i}}\omega t}{\rm{d}}t \notag\\
	      &\qquad\qquad+ \sum_{n}a_{n}b\int e^{{\rm{i}}\left[(n\omega_{\rm{SB}}-\Delta\omega)t + \phi_{n} - \phi_{\rm LO}\right]}e^{{\rm{-i}}\omega t}{\rm{d}}t
	      + \sum_{n}a_{n}b\int e^{{\rm{-i}}\left[(n\omega_{\rm{SB}}-\Delta\omega)t + \phi_{n} - \phi_{\rm LO}\right]}e^{{\rm{-i}}\omega t}{\rm{d}}t
	     \Big\} 
	       \label{l2}\\
	     &=  \frac{c\varepsilon}{2\sqrt{2\pi}}         
	    \Big\{
	     \left(b^2 + \sum_{n}a_{n}^2\right)\int e^{{\rm{-i}}\omega t}{\rm{d}}t
	     + \sum_{\substack{n,m\\ n\neq m}} a_{n}a_{m}e^{{\rm{i}}(\phi_{m} - \phi_{n})}\int e^{{\rm{i}}(m-n)\omega_{\rm{SB}}t}e^{{\rm{-i}}\omega t}{\rm{d}}t \notag\\
	      &\qquad\qquad+ \sum_{n}a_{n}be^{{\rm{i}} (\phi_{n} - \phi_{\rm LO})}\int e^{{\rm{i}}(n\omega_{\rm{SB}}-\Delta\omega)t }e^{{\rm{-i}}\omega t}{\rm{d}}t
	      + \sum_{n}a_{n}be^{{\rm{-i}}(\phi_{n} - \phi_{\rm LO})}\int e^{{\rm{-i}}(n\omega_{\rm{SB}}-\Delta\omega)t}e^{{\rm{-i}}\omega t}{\rm{d}}t
	     \Big\} 
	       \label{l3}	       
\end{align}

The Fourier integrals can be evaluated by noting that:
\begin{equation}
	\frac{1}{\sqrt{2\pi}}\int e^{{\rm i}\alpha t} e^{-{\rm i}\omega t} {\rm d}t= \sqrt{2\pi}\,\delta(\omega - \alpha),
	\label{m}
\end{equation}
where $\delta(\omega)$ is the Dirac delta function.

Substituting this in equation \ref{l2}, taking the corresponding value of $\alpha$ in each integrand gives:
\begin{align}
	\breve{I}_{\rm tot}(\omega) &=   \frac{c\varepsilon\sqrt{2\pi}}{2}         
	    \Big\{
	     \left(b^2 + \sum_{n}a_{n}^2\right)\delta(\omega) \notag\\
	     &\qquad\qquad+ \sum_{\substack{n,m\\ n\neq m}} a_{n}a_{m}e^{{\rm{i}}(\phi_{m} - \phi_{n})}\delta(\omega-[m-n]\omega_{\rm{SB}}) \notag\\
	      &\qquad\qquad+ \sum_{n}a_{n}b e^{{\rm{i}}(\phi_{n} - \phi_{\rm LO})}\delta(\omega-[n\omega_{\rm{SB}}-\Delta\omega]) \notag\\
	      &\qquad\qquad+ \sum_{n}a_{n}b e^{{\rm{-i}}(\phi_{n} - \phi_{\rm LO})}\delta(\omega+[n\omega_{\rm{SB}}-\Delta\omega])
	     \Big\} 
	       \label{n}
\end{align}

Equation \ref{n} describes three combs of teeth the in frequency domain, each with spacing $\omega_{SB}$, and a peak also at zero frequency representing the DC level. The two combs that sum only over $n$ represent actual optical frequencies present in the signal beam. There are two of these combs because each actual unique signal comb tooth has a positive and negative frequency component when the original electric field is represented in a form such as \ref{apxd}. 

The term that sums over $n$ and $m$ represents the beatnotes formed when the signal beam comb teeth beat against \emph{each other}. Different pairs of frequency components with equal frequency spacing will beat at the same frequency, and may add or subtract to that beatnote depending on their phase. For a set of sidebands produced by a phase only electro-optic modulator, the different phases between all frequency components add in a way so this ``cross comb" does not show up at all.

\begin{figure}
    \includegraphics[width=\columnwidth]{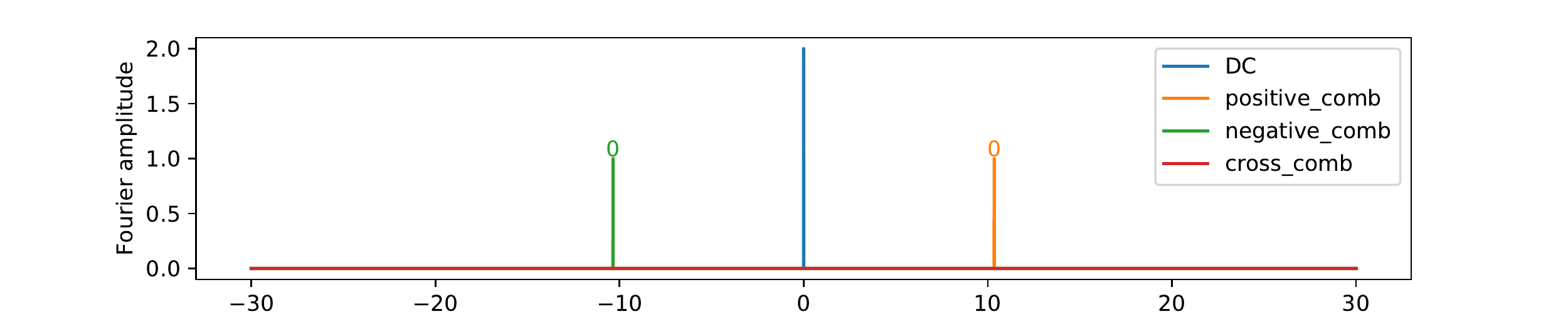}
    \caption{Beatnote from comb with single tooth.}
    \label{fig1}
\end{figure}

\begin{figure}
    \includegraphics[width=\columnwidth]{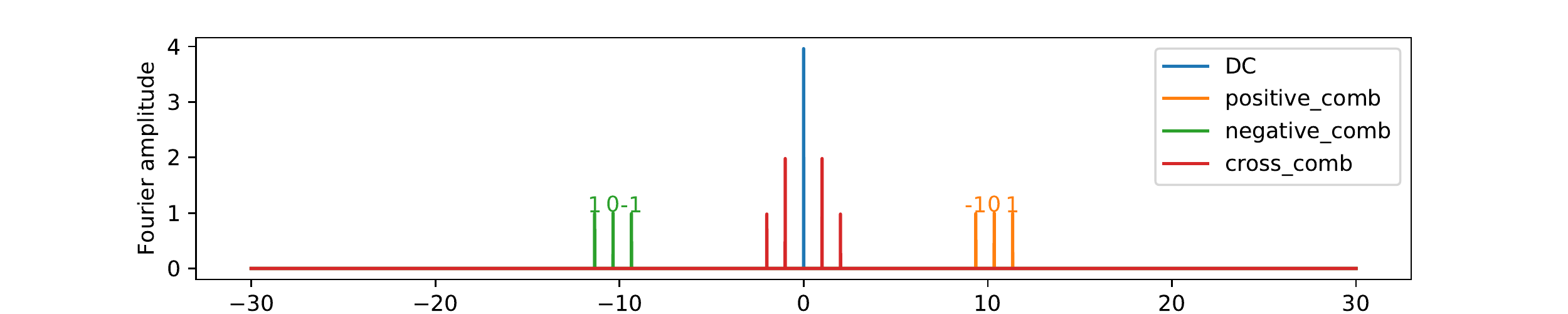}
    \caption{Beatnote from comb with 3 teeth.}
    \label{fig2}
\end{figure}

\begin{figure}
    \includegraphics[width=\columnwidth]{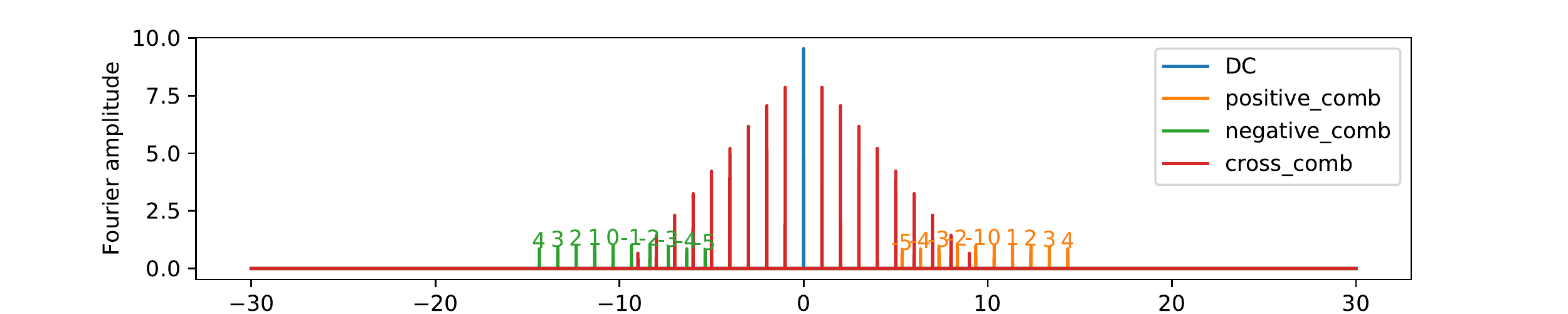}
    \caption{Beatnote from comb with 10 teeth.}
    \label{fig3}
\end{figure}

\begin{figure}
    \includegraphics[width=\columnwidth]{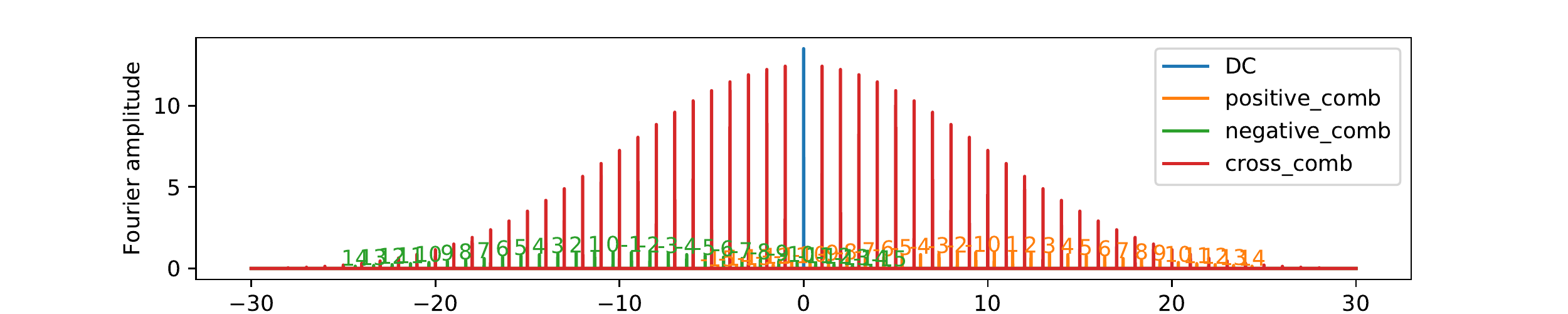}
    \caption{Beatnote from comb with 30 teeth.}
    \label{fig4}
\end{figure}

Depending on the number of teeth, the sideband spacing, and the local oscillator frequency offset, it is possible for teeth belonging to all three combs to overlap. It is generally undesirable for the teeth to overlap, since this prevents measurement of the amplitude and phase of individual of the individual frequency components that make up the signal beam frequency comb. Figures \ref{fig1}, \ref{fig2}, \ref{fig3} and \ref{fig4} show how combs can begin to intersect as the number of teeth get larger. Note that $\Delta\omega$ has deliberately been chosen to be an integer plus one third times the sideband spacing, so that even though all three combs become interspersed, the individual comb teeth do not lie on top of one another. The amplitude of the signal comb teeth was was set to be a Gaussian simply as a visual aid.

Equation \ref{n} gives the Fourier amplitude at a given frequency, but in reality, any measurement we make of the Fourier amplitudes is over some small range of frequencies. In a spectrum analyser, this range is set by the resolution bandwidth RBW, and in a numerical discrete Fourier transform of a time domain signal, it is set by the spacing of the frequency array (which is the inverse of the total length of time sampled). So experimentally, to extract the actual value for the Fourier amplitude at some frequency $\bar{\omega}$, we must integrate over some small frequency range : $\omega=\bar{\omega}\pm\epsilon$, and the resulting quantity will be denoted $\bar{\breve{I}}_{\rm tot}(\bar{\omega})$. Integration over the Dirac deltas means that the resulting values of $\bar{\breve{I}}_{\rm tot}(\bar{\omega})$ will be finite.

\begin{align}
	\bar{\breve{I}}_{\rm tot}(\bar{\omega}) &= \int_{\bar{\omega}-\epsilon}^{\bar{\omega}+\epsilon}  \breve{I}_{\rm tot}(\omega)  {\rm d}\omega  \label{o1} \\
	    &=\frac{c\varepsilon\sqrt{2\pi}}{2}         
	    \Big\{
	     \left(b^2 + \sum_{n}a_{n}^2\right)
	     \int_{\bar{\omega}-\epsilon}^{\bar{\omega}+\epsilon}\delta(\omega){\rm d}\omega \notag\\
	     &\qquad\qquad+ \sum_{\substack{n,m\\ n\neq m}} a_{n}a_{m}e^{{\rm{i}}(\phi_{m} - \phi_{n})}\int_{\bar{\omega}-\epsilon}^{\bar{\omega}+\epsilon}\delta(\omega-[m-n]\omega_{\rm{SB}}) {\rm d}\omega \notag\\
	      &\qquad\qquad+ \sum_{n}a_{n}b e^{{\rm{i}}(\phi_{n} - \phi_{\rm LO})}\int_{\bar{\omega}-\epsilon}^{\bar{\omega}+\epsilon}\delta(\omega-[n\omega_{\rm{SB}}-\Delta\omega]) {\rm d}\omega \notag\\
	      &\qquad\qquad+ \sum_{n}a_{n}b e^{{\rm{-i}}(\phi_{n} - \phi_{\rm LO})}\int_{\bar{\omega}-\epsilon}^{\bar{\omega}+\epsilon}\delta(\omega+[n\omega_{\rm{SB}}-\Delta\omega]) {\rm d}\omega
	     \Big\} 
	       \label{o2}
\end{align}

Equation \ref{o2} provides us with a simple method to determine the amplitude and phase of the optical signal comb teeth by measuring the radio frequency beat note. If the time domain radio signal is measured on an oscilloscope, it can be Fourier transformed numerically to show the radio-frequency combs. Assuming the teeth of the different combs are not overlapped then the Fourier amplitude of at the frequency of the $n$th tooth has a contribution from only a single term of equation \ref{o2}. Under these conditions, the optical amplitude and phase of the $n$th tooth is related to the Fourier transform of the detected intensity simply by:
\begin{equation}
	\bar{\breve{I}}_{\rm tot}(\bar{\omega}_n) = a_{n}b e^{{\rm{i}}(\phi_{n} - \phi_{\rm LO})},
	\label{p}
\end{equation}
so that the optical amplitude of the tooth is:
\begin{equation}
	a_{n} = \frac{|\bar{\breve{I}}_{\rm tot}(\bar{\omega}_n)|}{b},
	\label{q}
\end{equation}
and the optical phase of the tooth is:
\begin{equation}
	\phi_{n} = {\rm Arg}[\bar{\breve{I}}_{\rm tot}(\bar{\omega}_n)]  + \phi_{\rm LO}.
	\label{r}
\end{equation}

It is worth noting that the optical phase of the local oscillator $\phi_{\rm LO}$ will not usually be fixed in a heterodyne measurement. While it is certainly possible to set up some phase locking circuit by either feeding back to the laser or to a piezo controlled mirror, generally the $\phi_{\rm LO}$ will drift over time, as is seen in any unlocked interferometer. Therefore, the recovered value of $\phi_{n}$ will also drift over time. However, so long as $\phi_{\rm LO}$ stays constant over the time scale of a single measurement, then the \emph{difference} between all the $\phi_{n}$s will always be measured to be the same, even though $\phi_{\rm LO}$ may drift.

A final word about the beat note formed by the comb teeth beating against each: just because there is a comb, \emph{does not} mean you will actually observe any modulation of the intensity. Looking at the second term in equation \ref{n} (the term responsible for teeth beating with each other), it can be seen that pairs of $n,m$ with the same difference, do not necessarily have the same phase. If the phase of different pairs is not the same, then it is possible (indeed likely) that subsequent pairs will not all add in a way that increases the magnitude, and may totally cancel each other out. This is the case when a beam is phase modulated using an EOM. If the beam intensity is subsequently measured with only a photodiode, no intensity modulation is seen at all. However beating with a frequency shifted local oscillator reveals that there may be many other frequency components present.

\appendix
\section{Complex representation of a real wave}
The electric component of a electromagnetic wave propagating in the positive $x$ direction can be written as:
\begin{equation}
	E(x, t) = E_{0}(x, t)\cos{(kx-\omega t + \phi)},
	\label{apxa}
\end{equation}
where $E_{0}(x, t)$ is a real, possibly time varying amplitude, $k$ is the wave number, and $\phi$ is some constant phase offset.

The cosine can be written in complex exponential form:
\begin{align}
	E(x, t) &= \frac{E_{0}(x, t)}{2}\left(e^{{\rm{i}}(kx-\omega t + \phi)} + e^{-{\rm{i}}(kx-\omega t + \phi)}\right) \label{apxb1}\\
	       &= \frac{E_{0}(x, t)e^{{\rm{i}}(kx+ \phi)}}{2}e^{-{\rm{i}}\omega t} + \frac{E_{0}(x, t)e^{-{\rm{i}}(kx+ \phi)}}{2}e^{{\rm{i}}\omega t}.
	\label{apxb2}
\end{align}

Now, if we no longer care about the position dependence of the field, we can drop the explicit dependence on $x$, and simply note that the position would add some phase offset, in exactly the same way that $\phi$ does. The terms preceding the time dependent exponential can therefor just be written as a single complex valued amplitude $\mathcal{E}(t)$, noting that the two coefficients have the same magnitude, and negative phase (ie, they are the complex conjugate of one another):
\begin{align}
	\mathcal{E}(t) &= E_{0}(x, t)e^{{\rm{i}}(kx+ \phi)} \label{apxc1} \\
	\mathcal{E}^*(t) &= E_{0}(x, t)e^{-{\rm{i}}(kx+ \phi)}.
	\label{apxc}
\end{align}

Substitution into equation \ref{apxb2} yields out desired form for the electric field:
\begin{equation}
	E(t) = \frac{1}{2}\mathcal{E}(t)e^{-{\rm{i}}\omega t} + \frac{1}{2}\mathcal{E}^*(t)e^{{\rm{i}}\omega t}
	\label{apxd}
\end{equation}

Now, there are a couple of good questions about why we want the electric field represented in this form. Firstly, why didn't we just leave it with a cosine as in equation \ref{apxa} where everything was real? This is easy to answer: multiplication and integration etc. with complex exponentials is much, much easier (you don't need an encyclopaedia of trig identities in your head).
The second, more subtle question, is why don't we just work with the equation in the form of equation \ref{apxb1}, where the amplitude is real? The first part of the answer is that we don't want to keep writing the $kx + \phi$ part of the exponent, since we already said that we don't care about position dependence. The obvious next question is then: `if this just adds some constant phase, why don't you just set this to zero so that the amplitudes would be real?' In many cases you can, since you actually won't care about the phase, but you do loose generality by doing this. In order to keep the equation totally general, the coefficients ($\mathcal{E}$) are made to be complex so they can have any phase, and the relationship between the coefficients is enforced by observing one is the conjugate of the other. In many cases you might actually care about the position dependence, $kx$, or the initial phase $\phi$, and these absolutely can be left in remaining exponential term. The purpose then of the complex amplitude is that it can represent \emph{ANY} modulation of the reference wave, including additions of other phase or frequency components that you might wish to introduce, a property which is made use of in equation \ref{e}.

Finally, it is worth noting that there are two alternative (but equivalent) ways of writing equation \ref{apxd} that are in common use:
\begin{align}
	E(t) &= \frac{1}{2}\mathcal{E}(t)e^{-{\rm{i}}\omega t} + {\rm c.c.} \label{apxe1}\\
	       &= {\rm{Re}}\left\{ \mathcal{E}(t)e^{-{\rm{i}}\omega t} \right\} \label{apxe2}\\
	       &= |\mathcal{E}(t)|\cos{\Big(-\omega t + {\rm Arg}\big(\mathcal{E}(t)\big)\Big)}  \label{apxe3}
\end{align}
where $\rm{c.c.}$ is short for `complex conjugate', ${\rm{Re}}\{\}$ means to take the real part of the expression, and ${\rm Arg}$ returns the phase of a complex number. Generally, since it is known that the final value of the electric field \emph{must} be real, then any equation involving the field will \emph{always} contain the sum of two expressions which are complex conjugates of one another. Therefore, considerable time (and paper) can be saved by conducting \emph{all} calculations with only one of the pair of expressions, then at the end of the calculation, the full expression can be written out simply by adding on the conjugate. I personally like holding onto both terms, since when doing atomic physics, often approximations are made (like the rotating wave approximation), where certain terms are thrown away. Keeping track of all terms makes these approximations more intuitive and easier to follow in my opinion. Equation \ref{apxe3} is shown to make clear the relationship to a regular real cosine, and also to illustrate the purpose of putting factors of $1/2$ in forms like equation \ref{apxe1}, ie. including factors of $1/2$ makes the magnitude of the complex amplitude equal to the real amplitude.

%\bibliographystyle{pra}
%\bibliographystyle{iopart-num}

%\bibliographystyle{IEEEtran}
%\bibliography{heterodyne_refs}

\bibliography{heterodyne_derivation.bbl}

\begin{thebibliography}{1}
\providecommand{\url}[1]{#1}
\csname url@samestyle\endcsname
\providecommand{\newblock}{\relax}
\providecommand{\bibinfo}[2]{#2}
\providecommand{\BIBentrySTDinterwordspacing}{\spaceskip=0pt\relax}
\providecommand{\BIBentryALTinterwordstretchfactor}{4}
\providecommand{\BIBentryALTinterwordspacing}{\spaceskip=\fontdimen2\font plus
\BIBentryALTinterwordstretchfactor\fontdimen3\font minus
  \fontdimen4\font\relax}
\providecommand{\BIBforeignlanguage}[2]{{%
\expandafter\ifx\csname l@#1\endcsname\relax
\typeout{** WARNING: IEEEtran.bst: No hyphenation pattern has been}%
\typeout{** loaded for the language `#1'. Using the pattern for}%
\typeout{** the default language instead.}%
\else
\language=\csname l@#1\endcsname
\fi
#2}}
\providecommand{\BIBdecl}{\relax}
\BIBdecl

\bibitem{kuri2003optical}
T.~Kuri and K.-i. Kitayama, ``Optical heterodyne detection technique for
  densely multiplexed millimeter-wave-band radio-on-fiber systems,''
  \emph{Journal of lightwave technology}, vol.~21, no.~12, pp. 3167--3179,
  2003.

\bibitem{delange1968optical}
O.~DeLange, ``Optical heterodyne detection,'' \emph{IEEE spectrum}, vol.~5,
  no.~10, pp. 77--85, 1968.

\bibitem{chtcherbakov2007optical}
A.~A. Chtcherbakov, R.~J. Kisch, J.~D. Bull, and N.~A. Jaeger, ``Optical
  heterodyne method for amplitude and phase response measurements for
  ultrawideband electrooptic modulators,'' \emph{IEEE photonics technology
  letters}, vol.~19, no.~1, pp. 18--20, 2007.

\bibitem{hall1981optical}
J.~Hall, L.~Hollberg, T.~Baer, and H.~Robinson, ``Optical heterodyne saturation
  spectroscopy,'' \emph{Applied Physics Letters}, vol.~39, no.~9, pp. 680--682,
  1981.

\bibitem{jacobs1988optical}
S.~Jacobs, ``Optical heterodyne (coherent) detection,'' \emph{American Journal
  of Physics}, vol.~56, no.~3, pp. 235--245, 1988.

\bibitem{keyes2013optical}
R.~Keyes, \emph{Optical and Infrared Detectors}, ser. Topics in Applied
  Physics.\hskip 1em plus 0.5em minus 0.4em\relax Springer Berlin Heidelberg,
  2013.

\bibitem{menzies2005laser}
R.~T. Menzies, ``Laser heterodyne detection techniques,'' \emph{Laser
  Monitoring of the Atmosphere}, pp. 297--353, 2005.

\end{thebibliography}

\end{document}